\documentclass[paper]{JHEP3}

\preprint{KEK-TH-1141}

\usepackage{epsfig,multicol}
\usepackage{epsfig}
\usepackage{amsmath}
\usepackage{amssymb}

\title{
Monte Carlo approach to nonperturbative strings\\
--- demonstration in 
noncritical string theory
}

\author{ 
Naoyuki Kawahara${}^{a}$\,, Jun Nishimura${}^{ab}$ 
and Atsushi Yamaguchi${}^{a}$ 
\vspace*{0.5cm} \\ 
\llap{$^a$}High Energy Accelerator Research Organization (KEK),\\
1-1 Oho, Tsukuba 305-0801, Japan  \\
\llap{$^b$}Department of Particle and Nuclear Physics,\\
Graduate University for Advanced Studies (SOKENDAI),\\
1-1 Oho, Tsukuba 305-0801, Japan 
\vspace*{0.5cm} \\ 
\email{kawahara@post.kek.jp,
jnishi@post.kek.jp, ayamagu@post.kek.jp}
}

\abstract{
We show how Monte Carlo approach can be used to 
study the double scaling limit in matrix models.
As an example, we study a solvable hermitian one-matrix model
with the double-well potential, which
has been identified recently as a dual description
of noncritical string theory
with worldsheet supersymmetry.
This identification utilizes the nonperturbatively
stable vacuum unlike its bosonic counterparts,
and therefore it 
provides a complete constructive formulation 
of string theory.
Our data with the matrix size ranging from 
$8$ to $512$
show a clear scaling behavior, which enables
us to extract the double scaling limit accurately.
The ``specific heat'' obtained in this way agrees nicely with 
the known result obtained
by solving the Painleve-II equation
with appropriate boundary conditions.
}

\def\be{\begin{equation}}
\def\ee{\end{equation}}
\def\beq{\begin{equation}}
\def\eeq{\end{equation}}
\def\bear{\begin{eqnarray}}
\def\eear{\end{eqnarray}}
\def\beqa{\begin{eqnarray}}
\def\eeqa{\end{eqnarray}}

\newcommand{\tr}{{\rm tr\,}}
\newcommand{\Range}{r}

\begin{document}
\section{Introduction}

Matrix models have been considered as one of the most
powerful frameworks to formulate string theories
in a nonperturbative manner.
A fundamental viewpoint which links matrix models
to string theories was given by 't Hooft \cite{'tHooft:1973jz}.
There Feynman diagrams which appear in matrix models are
identified with {\em discretized} string worldsheets.
However, if one takes the large-$N$ limit naively
(the so-called planar limit),
only the planar diagrams survive, 
which implies the appearance of a classical string theory.
One way to formulate nonperturbative string theory
using matrix models is therefore 
to look for a nontrivial large-$N$ limit, 
in which Feynman diagrams with all kinds of 
topology survive.\footnote{
Another possibility to realize string theory using matrix models
is to keep $N$ finite
as in the AdS/CFT correspondence \cite{Aharony:1999ti}, 
topological string theory \cite{Gopakumar:1998ki_Dijkgraaf:2002fc}
and the Kontsevich model \cite{Kontsevich:1992ti}.}
The existence of such a limit
has been first demonstrated in matrix models 
for noncritical string theory
\cite{Brezin:1990rb,Douglas:1989ve,Gross:1989vs},
and it is called the double scaling limit.
(See also refs.\ 
\cite{DSLgeneral,Bietenholz:2002ch,Bietenholz:2004xs,Bietenholz:2006cz}
for recent works, in which
the double scaling limit appears in various contexts.)

It is generally believed that a similar idea can be 
applied also to critical string theories.
The corresponding matrix models have been proposed
in refs.\ \cite{BFSS,Ishibashi:1996xs,DVV},
but the existence of a nontrivial large-$N$ limit
is yet to be confirmed.
%
%
To address such an issue, the 2d Eguchi-Kawai model \cite{EK}
has been studied as a toy model.
Indeed Monte Carlo simulation \cite{Nakajima:1998vj} demonstrated
the existence of
a one-parameter family of large-$N$ limits, which 
generalizes the Gross-Witten 
\cite{Gross:1980he} planar large-$N$ limit.
If one modifies the Eguchi-Kawai model by introducing the 
twist \cite{GAO},
the double scaling limit
can be identified with the continuum limit of
field theories on discrete non-commutative (NC) geometry \cite{AMNS}.
The actual existence of such limits 
has been demonstrated by
Monte Carlo simulations
in the case of NC gauge theory in 2d \cite{Bietenholz:2002ch}
and 4d \cite{Bietenholz:2006cz} and also
in 3d NC scalar field theory \cite{Bietenholz:2004xs}.
In all these cases, it was observed
that non-planar diagrams indeed affect
the infrared dynamics drastically through the
UV/IR mixing mechanism \cite{MRS}.

We consider that
Monte Carlo simulation would be a powerful tool
also to study matrix models for critical string theories.
%
%
Technically the IIB matrix model \cite{Ishibashi:1996xs} would be the
least difficult among them since the space-time, on which
the ten-dimensional ${\cal N}=1$ super Yang-Mills theory is defined,
is totally reduced to a point.
However, the integration over the fermionic matrices yields
a complex Pfaffian, which makes the Monte Carlo simulation
still very hard \cite{Ambjorn:2000dx,Anagnostopoulos:2001yb}.
An analogous model, which can be obtained by dimensionally
reducing {\em four-dimensional} ${\cal N}=1$ 
super Yang-Mills theory to a point, 
does not have that problem, and Monte Carlo studies 
suggest the existence of a nontrivial large-$N$ limit 
\cite{Ambjorn:2000bf}.

The developments in the matrix description of
critical string theories have also given a new perspective
to noncritical string theory.
%
For instance, in matrix quantum mechanics which describe
$(1+1)$-dimensional string theory in the double scaling limit,
the matrix degrees of freedom
have been interpreted as the tachyonic open-string field 
living on unstable D0-branes \cite{McGreevy:2003kb,Klebanov:2003km}.
Based on this interpretation,
matrix models with the double-well potential, which are
known to be solvable, have been identified as a dual description of  
noncritical string theory with worldsheet supersymmetry
\cite{Takayanagi:2003sm,Douglas:2003up,Klebanov:2003wg}.
An important property of these models is 
that they possess a stable nonperturbative vacuum
unlike their bosonic counterparts,
and therefore one can obtain a complete constructive formulation
of string theory.
It also provides us with a unique opportunity to test the
validity and the feasibility of Monte Carlo methods for studying 
string theories nonperturbatively.
In particular we are concerned with such questions as
what kind of analysis is possible
to extract the double scaling limit,
and how large the matrix size should be.

In this work we consider the simplest model \cite{Cicuta:1986pu},
namely a hermitian one-matrix model
identified \cite{Klebanov:2003wg}
as a dual of $\hat c=0$ noncritical string theory,\footnote{The 
model studied in this paper
was also used in ref.\ \cite{Kawai:2004pj} 
to calculate the chemical potential of D-instantons,
which is shown to be a universal quantity in the
double scaling limit \cite{Hanada:2004im}.   
These works are generalized to 
other noncritical string theories \cite{Sato:2004tz,%
Ishibashi:2005zf,Matsuo:2005nw}
and discussed in various contexts 
\cite{deMelloKoch:2004en,Ishibashi:2005dh,%
Fukuma:2005nm,Kuroki:2007an}.}
which is sometimes referred to as the pure supergravity 
in the literature.
We calculate correlation functions 
near the critical point, and investigate their scaling behavior
to extract the double scaling limit. The results are then
compared with a prediction obtained by a different approach.
We hope that the lessons from this work
would be useful in applying the same method to models
which are not accessible by analytic methods.




The rest of this paper is organized as follows.
In section \ref{section:MA} we introduce the one-matrix model, 
and present some simulation details.
In section \ref{section:PL} we obtain explicit results
in the planar limit,
and compare them with the known analytical results.
In section \ref{section:DSL} we search for
a double scaling limit by using only Monte Carlo data.
The results are compared with the prediction obtained
by the orthogonal-polynomial technique. 
In section \ref{section:exact} we present more
detailed comparison with the analytical prediction.
Section \ref{section:summary} is devoted to a summary and discussions.
In the Appendix we briefly review the derivation of some asymptotic
behaviors in the double scaling limit.

\section{The model and some simulation details}
\label{section:MA}

The model we study in this paper is defined by
\bear
Z&=& \int d^{N^2} \!\!  \phi \, 
\exp\left( -S \right) \ ,
\label{Zdef}
\\
S &=& \frac{N}{g} \, \tr\left(-\phi^2 + \frac{1}{4}\phi^4 \right) \ ,
\label{omm}
\eear 
where $\phi$ is an $N\times N$ hermitian matrix.
We assume that the coupling constant $g$ is positive
so that the action is bounded from below.
Since the action takes the form of a double-well,
the standard Metropolis algorithm using a trial configuration
obtained by slightly modifying some components of the matrix 
would have a problem with ergodicity.
In order to circumvent this problem, we perform the simulation
as follows.

Let us diagonalize the hermitian matrix $\phi$ as
$\phi = U \Lambda U^{-1}$,
where $\Lambda = {\rm diag}(\lambda_1,\cdots,\lambda_N)$
is a real diagonal matrix.
Due to the SU($N$) invariance of the model,
the angular variable $U$ can be integrated out.\footnote{An analogous 
model including a kinetic term
representing the fuzzy sphere background has been
studied by Monte Carlo simulation in refs.\ \cite{Martin,Panero}.
The basic idea to avoid the ergodicity
problem can be applied there as well,
although in that case the angular variables $U$ have to be treated
in Monte Carlo simulation. We thank 
Marco Panero for communications on this issue.}
Thus we are left with a system of eigenvalues $\lambda_i$
\bear
Z &=& \int \prod_{i=1}^{N} 
d\lambda_i \, \exp (-\widetilde{S}) \ , 
\label{d_pf} \\
\widetilde{S} &=& \frac{N}{g}
\sum_{i=1}^{N}\left(-\lambda_i^2 + \frac{1}{4}\lambda_i^4
\right)
- \sum_{i < j}  \log |\lambda_i-\lambda_j|^2 \ ,
\label{d_action}
\eear
where the log term in eq.\ (\ref{d_action})
comes from the Vandermonde determinant. 
Due to the $\lambda _i ^4$ term in the action,
the probability of $\lambda_i$ having a large absolute
value is strongly suppressed.
This can be seen also from
the eigenvalue distribution in fig.\ \ref{fig_rho},
which actually has a compact support in the 
planar large-$N$ limit; see eqs.\ (\ref{rho_largeN-1})
and (\ref{rho_largeN-2}).
We therefore restrict $|\lambda_i|$ to be less than 
some value $X$.

We first run a simulation with a reasonably large $X$.
By measuring the eigenvalue distribution, we can
obtain an estimate on $X$ that 
can be used without affecting the Monte Carlo results.
We generate a trial configuration by replacing
one eigenvalue by a uniform random number 
within the range $[-X ,X]$. 
The trial configuration is accepted as a new configuration
with the probability max($1,\exp(-\Delta \widetilde{S})$), 
where $\Delta \widetilde{S}$ 
is the increase of the action $\widetilde{S}$
($\Delta \widetilde{S}<0$ in case it decreases).
The acceptance rate turns out to be 
of the order of a few percent.\footnote{We could have
increased the acceptance rate by suggesting a number
for the eigenvalue with a non-uniform probability
and taking it into account in the Metropolis accept/reject
procedure. In this work, however, we stayed with the simplest
algorithm for illustrative purposes.}
We repeat this procedure for all the eigenvalues,
and that defines our ``one sweep''.

Typically we make 500,000 sweeps for each set of parameters.
We discard the first 10,000 sweeps for thermalization,
and measure quantities every 100 sweeps considering
auto-correlation. The statistical errors are estimated by
the standard jack-knife method, although in most cases 
the error bars are invisible compared with the symbol size.
The simulation has been performed on
PCs with Pentium 4 (3GHz), and it took a few weeks
to get results for each value of $g$ with the largest 
system size $N=2048$.
Note that the required CPU time is of O($N^2$) 
thanks to the fact that we only have to deal with
the eigenvalues but not the whole matrix degrees
of freedom. Otherwise the required CPU time would
grow as O($N^3$) at least.
Note also that our algorithm allows
the eigenvalues to move from one well
to the other with finite probability.
Thus the problem with ergodicity is 
avoided.

\section{The planar limit} 
\label{section:PL}

In this section we investigate
the planar limit of the model by Monte Carlo simulation.
This limit corresponds to sending the matrix size
$N$ to infinity with fixed $g$.
It is necessary to study the planar limit first since
we have to identify the critical point, and calculate
correlation functions at that point,
which will be used when we search for 
a double scaling limit.

\FIGURE[t]{
\epsfig{file=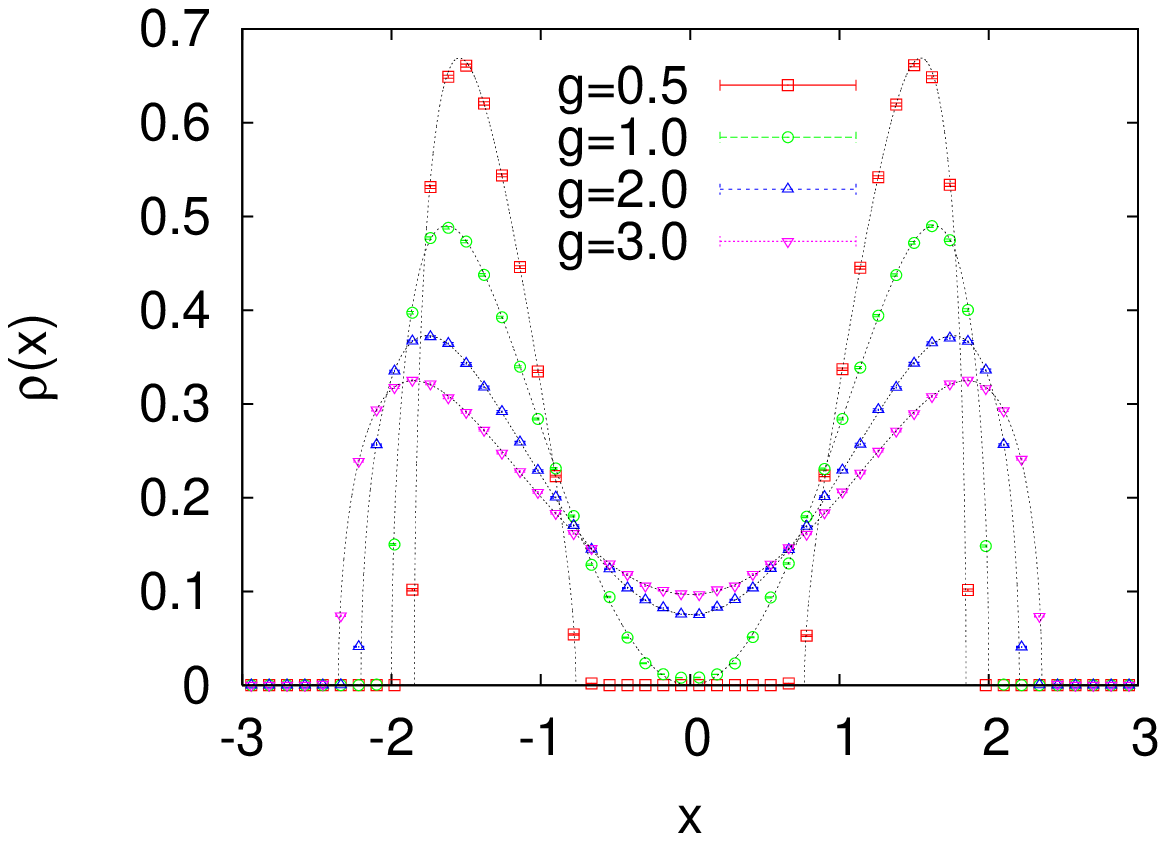, width=.5\textwidth}
\caption{The eigenvalue distribution $\rho(x)$ is plotted
for $g=0.5,1.0,2.0,3.0$ with $N=32$. 
The curves represent the exact results 
(\ref{rho_largeN-1}), (\ref{rho_largeN-2}) 
obtained in the planar large-$N$ limit.
}\label{fig_rho}}

Let us define the eigenvalue density distribution
\be
\rho(x)\equiv \frac{1}{N}
\Bigl\langle \tr\delta(x-\phi)
\Bigr\rangle \ ,
\label{rho}
\ee
from which one can calculate
the expectation value of any single trace operator.
In the planar limit the distribution $\rho(x)$ is 
obtained analytically \cite{Cicuta:1986pu}
using the method developed in ref.\ \cite{Brezin:1977sv}.
For $g\ge 1$ the distribution is given by
\beqa
\lim_{N\to \infty}\rho(x) &=&
\frac{1}{\pi g} \left(\frac{1}{2}x^2+\Range^2-1 \right)
\nonumber \\
&~& 
 \sqrt{4\Range^2-x^2} 
\label{rho_largeN-1}
\eeqa
in the range $-2\Range \leq x\leq 2\Range$,
where $\Range^2=\frac{1}{3}(1+\sqrt{1+3 g})$.
For $g \le 1$ it is given by
\be
\lim_{N\to \infty}\rho(x)=
\frac{1}{2\pi g}|x|\sqrt{(x^2-\Range_-^2)(\Range_+^2-x^2)}  
\label{rho_largeN-2}
\ee
in the range $ \Range_-\leq |x|\leq \Range_+ $, 
where $\Range_\pm^2=2(1\pm \sqrt{g})$.
Outside the specified region, 
the distribution is constantly zero,
and hence it has a compact support
for $g \ge 1$, which splits into two for $g \le 1$.
This implies a phase transition of the Gross-Witten type
\cite{Gross:1980he} at the critical point
\be
g=g_{\rm cr}\equiv 1 \ .
\ee 
Our Monte Carlo results for $N=32$
shown in fig.\ \ref{fig_rho}
agree well with the exact results 
in the planar limit.

\DOUBLEFIGURE[h] 
{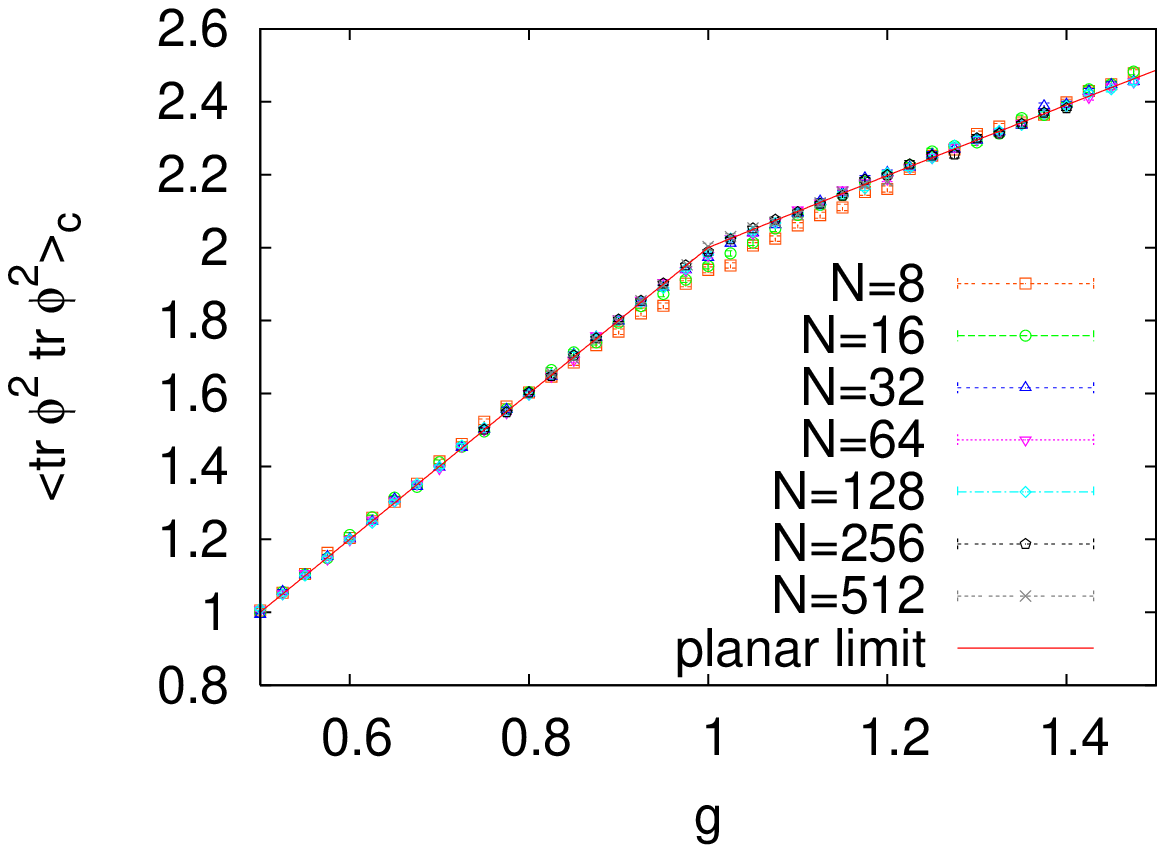, width=.49\textwidth}
{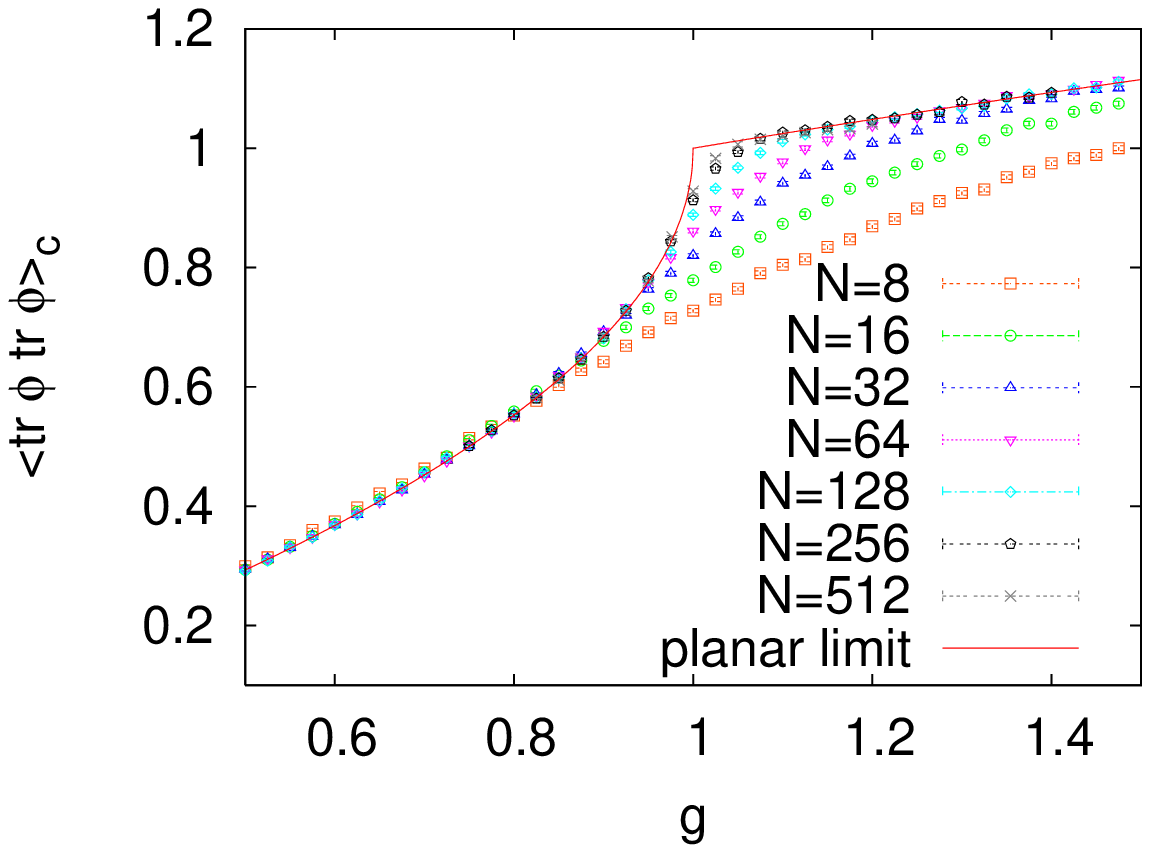, width=.49\textwidth}
{The two-point correlation function 
$\langle \tr \phi^2 \, \tr \phi^2 \rangle_{\rm c}$  
is plotted against $g$ for various $N$.
The solid line represents the analytic
result (\ref{p2p2_largeN}) in the planar large-$N$ limit.
\label{fig_p2p2}
}
{The two-point correlation function 
$\langle \tr \phi \, \tr \phi \rangle_{\rm c}$  
is plotted against $g$ for various $N$.
The solid line represents the analytic
result (\ref{pp_largeN}) in the planar large-$N$ limit.
 \label{fig_pp}}

Let us next consider two-point correlation functions
$\langle \tr\phi^2 \, \tr\phi^2 \rangle_{\rm c}$ and  
$\langle \tr\phi \, \tr\phi \rangle_{\rm c}$, 
where the suffix ``c'' 
implies that the connected part is taken.
In the planar limit the correlation functions are
obtained analytically as (See Appendix for derivation)
\bear
\lim_{N\to \infty}\langle\tr\phi^2 \, \tr\phi^2\rangle_{\rm c}
&=& \left\{
\begin{array}{cc}
\frac{2}{9}\left(1+\sqrt{1+3 g}\right)^2 
\quad & \textrm{for~} g \geq 1\ , \\
2g
\quad & \textrm{for~} g \leq 1\ . \\
\end{array}
\right.
\label{p2p2_largeN}\\
\lim_{N\to \infty}\langle\tr\phi \, \tr\phi\rangle_{\rm c}
&=& \left\{
\begin{array}{cc}
\frac{1}{3}\left(1+\sqrt{1+3 g}\right) 
\quad & \textrm{for~} g \geq 1 \ , \\
1-\sqrt{1-g}
\quad & \textrm{for~} g \leq 1 \ . \\
\end{array}
\right.\label{pp_largeN}
\eear
Our Monte Carlo results for various $N$ shown in
figs.\ \ref{fig_p2p2} and \ref{fig_pp} approach
the planar limit with increasing $N$.

In passing, let us consider 
the free energy of the system (\ref{Zdef}) defined by
\be
F\equiv \log Z- \frac{1}{4} N^2 \log g \ ,
\label{defF}
\ee
where the log term is subtracted in order to make
$F$ finite in the free case ($g=0$).
One can easily see that the correlation function
$\langle \tr\phi^2 \, \tr\phi^2 \rangle_{\rm c}$
is related to the second derivative of the free energy
with respect to $g^{-1/2}$ as
\be
\langle \tr\phi^2 \, \tr\phi^2 \rangle_{\rm c}
=\frac{g}{N^2} \, \frac{\partial^2
}{\partial(g^{-1/2})^2} \, F \ . 
\label{u_to_F}
\ee 
Therefore, the behavior (\ref{p2p2_largeN})
at the critical point $g=1$ 
implies that the phase transition is of third order
in accord with ref.\ \cite{Gross:1980he}.


\section{The double scaling limit}
\label{section:DSL}

In this section we search for 
a double scaling limit, in which 
we send the coupling constant
$g$ to the critical point $g_{\rm cr}= 1$
simultaneously with the $N \rightarrow \infty$ limit
keeping
\be
 \mu\equiv N^{p/3}(1
-g)
\label{mu}
\ee
fixed.
%
We investigate whether the quantities
\beqa
A (\mu , N) &\equiv&
- N^{q/3} \Bigl( \langle\tr\phi^2 \, 
\tr\phi^2\rangle_{\rm c}  - 2
\Bigr) \ , 
\label{p2p2_extracted}\\
B (\mu , N) &\equiv&
- N^{r/3}
\Bigl( \langle\tr\phi \, \tr\phi \rangle_{\rm c} - 1
\Bigr)
\label{pp_extracted}
\eeqa
have large-$N$ limits as functions of $\mu$
for some choice of the parameters $p$, $q$ and $r$.
In eqs.\ (\ref{p2p2_extracted}) and (\ref{pp_extracted}),
we have subtracted the values
in the planar large-$N$ limit
at the critical point $g=1$,
which are 2 and 1, respectively, for each correlation function;
see, eqs.\ (\ref{p2p2_largeN}) and (\ref{pp_largeN}).


\DOUBLEFIGURE[htb]
{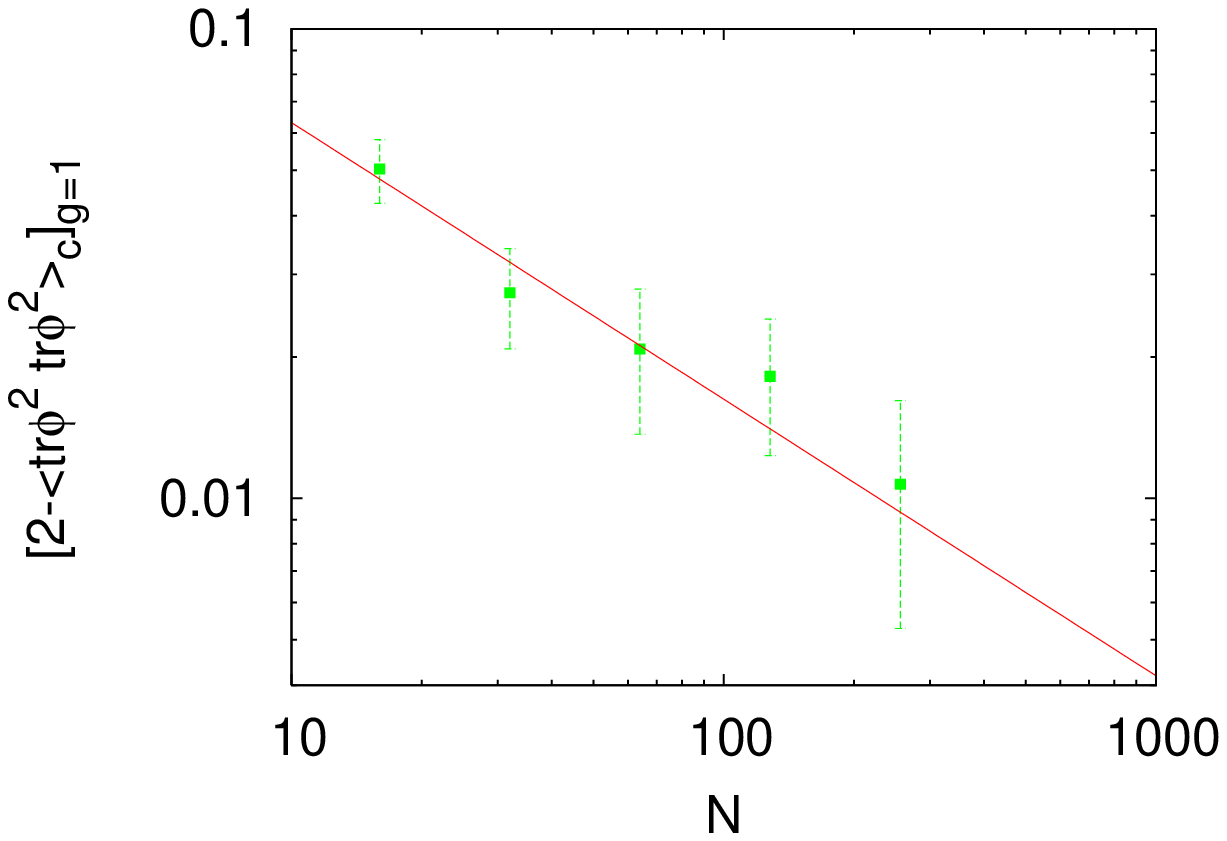, width=.49\textwidth}
{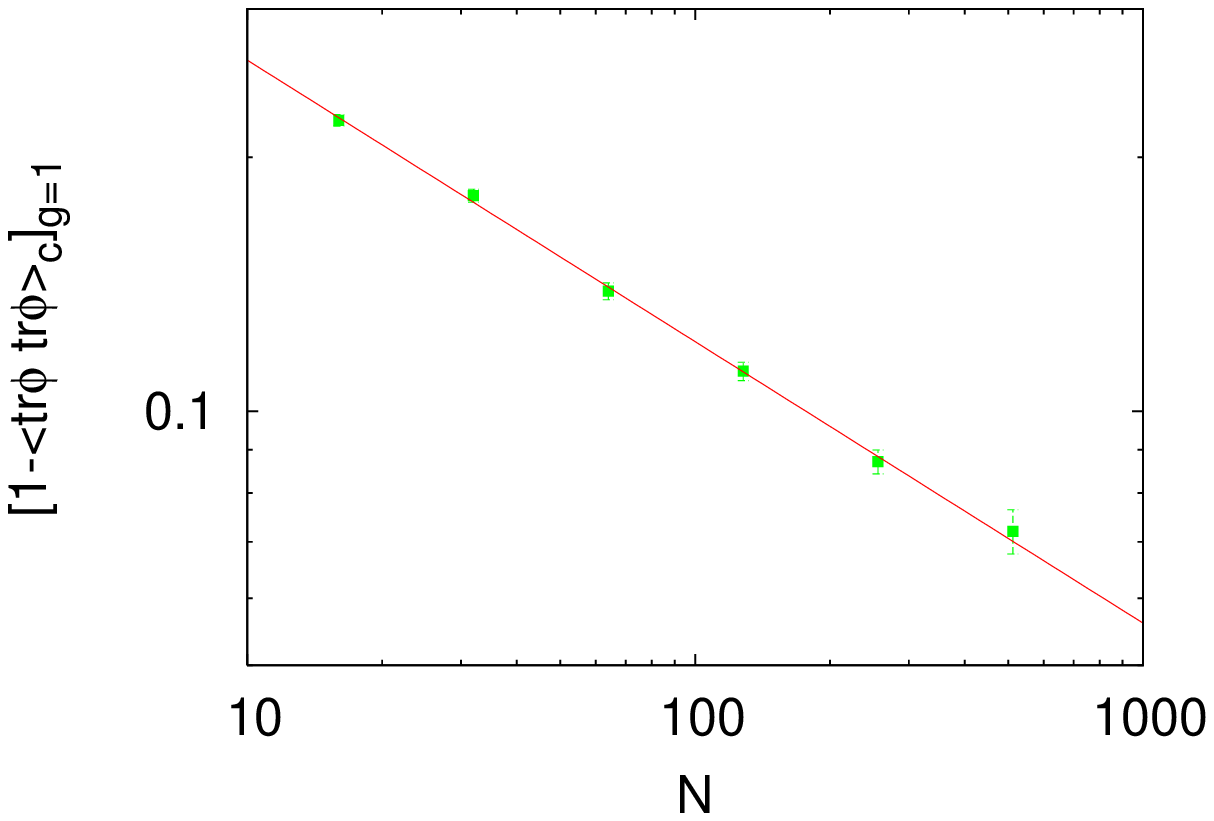, width=.49\textwidth}
{The observable $2-\langle \tr \phi^2 \, \tr \phi^2 \rangle_{\rm c}$ 
at the critical point $g=1$
is plotted against $N$ in the log-log scale.
The straight line represents a fit to the power law 
behavior $N^{-1.7(3)}$.
\label{mu0_graph1}}
{The observable $1-\langle \tr \phi \, \tr \phi \rangle_{\rm c}$ 
at the critical point $g=1$
is plotted against $N$ in the log-log scale.
The straight line represents a fit to the power law 
behavior $N^{-1.00(2)}$.
\label{mu0_graph2}}

In fact, by merely 
looking at the behavior of the planar results
(\ref{p2p2_largeN}) and (\ref{pp_largeN})
near the critical point $g\sim 1$,
one can readily deduce the existence of a double scaling limit
for $\mu \sim \pm \infty$,
where the $\pm$ sign corresponds 
to the behavior for $g \rightarrow 1 \pm \epsilon$,
respectively.
Namely, plugging $g = 1 - \mu N^{-p/3}$ into 
(\ref{p2p2_largeN}) and (\ref{pp_largeN}), one obtains
\beqa
\lim _{N\to \infty}
A(\mu, N) &=& 
\left\{
\begin{array}{cc}
2 \mu \quad & \textrm{for~} \mu\sim \infty \ , \\
\mu \quad & \textrm{for~} \mu\sim -\infty \ ,
\end{array}
\right.
\label{AtoNlim}
\\
\lim _{N\to \infty}
B(\mu, N) &=& 
\left\{
\begin{array}{cc}
\sqrt{\mu} \quad & \textrm{for~} \mu\sim \infty \ , \\
0  \quad & \textrm{for~} \mu\sim -\infty \ ,
\end{array}
\right.
\label{BtoNlim}
\eeqa
\beq
\mbox{~~with~~}
q=p  \quad {\rm and} \quad r=\frac{p}{2} \ .
\label{qr-p-rel}
\eeq

When we search for a double scaling limit,
we have to impose (\ref{qr-p-rel}) in order to ensure
the scaling behavior at large $|\mu|$.
The nontrivial question then is whether we can 
choose the parameters within the constraints (\ref{qr-p-rel})
in such a way that
the scaling extends to small $|\mu|$.
In general, the planar results can be used in this way to
impose some constraints on the parameters
that appear in searching for a double scaling limit.
A similar strategy has been used,
for instance,
in ref.\ \cite{Nakajima:1998vj,Bietenholz:2002ch,Bietenholz:2004xs}.
We emphasize, however, that 
this is just meant to make the analysis simpler,
and that the relation (\ref{qr-p-rel}) would come out
anyway when we attempt to optimize the scaling behavior 
at large $|\mu|$.
%
%

Let us search for a scaling behavior 
at the particular point $\mu=0$.
This corresponds to $g=1$ for any choice of $p$
due to (\ref{mu}), and therefore, we can actually
determine $q$ and $r$ without using (\ref{qr-p-rel}).
In fig.\ \ref{mu0_graph1} we plot
the r.h.s.\ of (\ref{p2p2_extracted}) omitting the 
factor $N^{q/3}$. The observed power behavior 
implies $q=1.7(3)$.
Similarly from fig.\ \ref{mu0_graph2}, 
we obtain $r=1.00(2)$.
Using this value of $r$, the other exponents
$p$ and $q$ may be obtained 
from the relation (\ref{qr-p-rel})
as $p=q=2.00(4)$.
This is consistent with the value of $q$ 
extracted from fig.\ \ref{mu0_graph1} directly.
The latter has a larger error bar, though.
The reason for this is that the quantities
plotted in figs.\ \ref{mu0_graph1} and \ref{mu0_graph2}
are of the order of $\frac{1}{N^2}$ and $\frac{1}{N}$,
respectively.

\DOUBLEFIGURE[htb]
{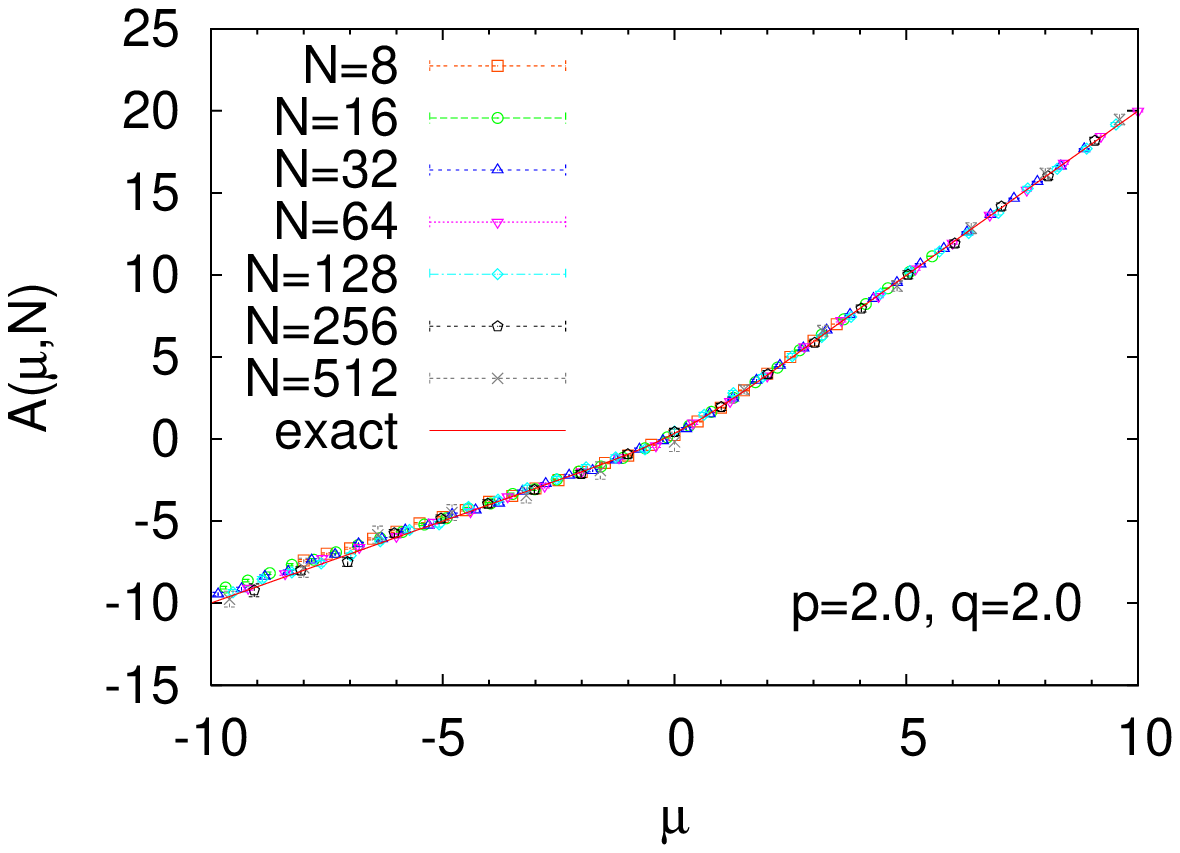,width=.49\textwidth}
{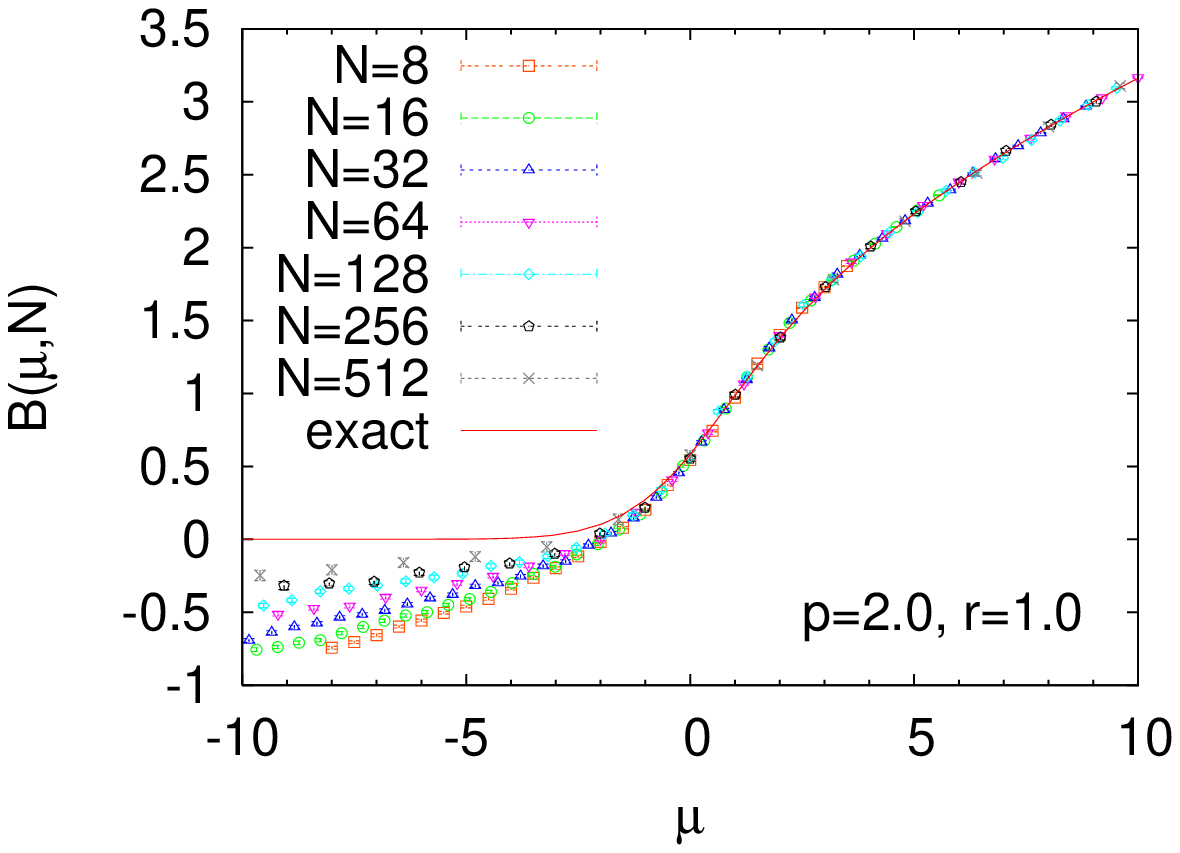, width=.49\textwidth}
{The quantity $A(\mu , N)$
is plotted against $\mu$ for $N=8, 16, \cdots , 512$
with $p=q=2.0$. 
The solid line represents the result (\ref{AtoN}) 
obtained in the double scaling limit.
\label{dsl_p2p2}}
{The quantity $B(\mu , N)$
is plotted against $\mu$ for $N=8, 16, \cdots , 512$
with $p=2.0$ and $r=1.0$.
The solid line represents the result (\ref{BtoN}) 
obtained in the double scaling limit.
\label{dsl_pp}}


Now let us see whether 
these values of $p$, $q$ and $r$
make the quantities $A (\mu , N)$ 
and $B(\mu , N)$ 
scale also for $\mu \neq 0$.
Using the Monte Carlo data shown in fig.\ \ref{fig_p2p2},
we plot the quantities as functions of $\mu$.
Figs.\ \ref{dsl_p2p2} and \ref{dsl_pp} show the results.
The scaling functions given below in eqs.\
(\ref{AtoN}) and (\ref{BtoN}) 
are also plotted for comparison. 
The Monte Carlo results for 
$A (\mu , N)$ 
show a nice scaling behavior, and they agree with
the prediction (\ref{AtoN}).
On the other hand, 
the quantity $B (\mu , N)$ scales and agrees
with the prediction (\ref{BtoN}) only in 
the $\mu \gtrsim 0$ region.
In the $\mu \lesssim 0$ region,
we observe 
some tendency towards scaling as $N$ increases up to $N=512$, 
but the convergence to the prediction (\ref{BtoN}) seems to be slow.
This behavior 
is due to the next-leading $1/N$ corrections,
as we discuss in the next section.

In fact the analysis based on the orthogonal polynomial technique
\cite{Brezin:1990rb,Douglas:1989ve,Gross:1989vs}
suggests the existence of a double scaling limit with 
\beq
p=q=2 \quad {\rm and} \quad r=1 \ ,
\label{q-r-p-sol}
\eeq
which agrees with our observation.
In this limit the model (\ref{omm})
is conjectured \cite{Klebanov:2003wg}
to be a dual description of the $\hat{c}=0$ 
noncritical string theory, where
the parameter $\mu$ is identified with the
cosmological constant in the corresponding
super Liouville theory. 
Note that we are able to deduce the existence of the 
double scaling limit only from Monte Carlo data.

Let us also note that due to eq.\ (\ref{u_to_F}),
$A (\mu , N)$ is related to the ``specific heat''
\beq
C(\mu , N ) \equiv
\frac{\partial^2 F}{\partial \mu^2} - 
 \left.
\frac{\partial^2 F}{\partial \mu^2} 
\right|_{\mu =0} 
\label{Cdef}
\eeq
as
\beq
A (\mu , N) = 
- \frac{1}{4} \, N^{(q+2p-6)/3}
\left\{
C(\mu , N ) + {\rm O}\left(N^{-2p/3}\right) \right\} \ .
\eeq
Therefore, the scaling of $A(\mu , N)$ with
the choice (\ref{q-r-p-sol}) 
implies that the ``specific heat'',
which has a physical meaning in the dual string theory,
becomes finite in the double scaling limit.
%

\DOUBLEFIGURE[htb]
{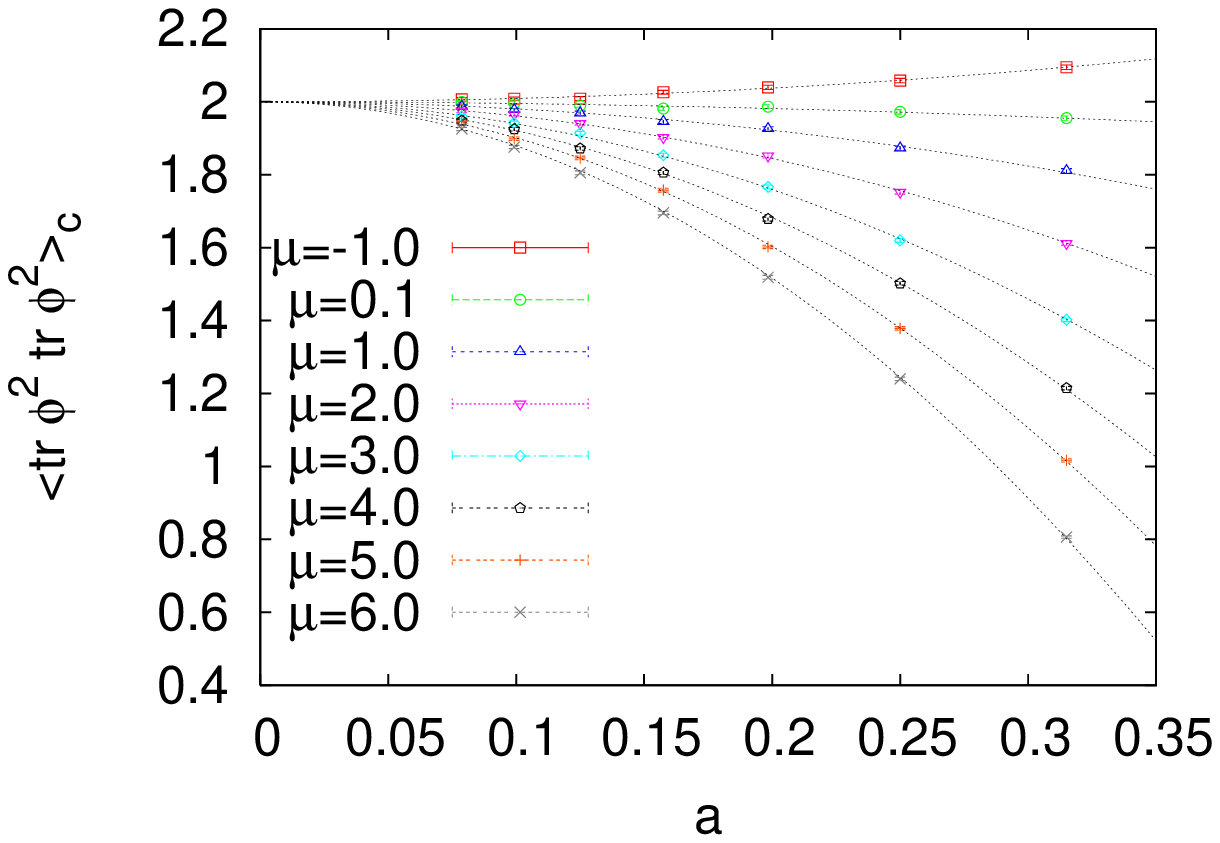, width=.49\textwidth}
{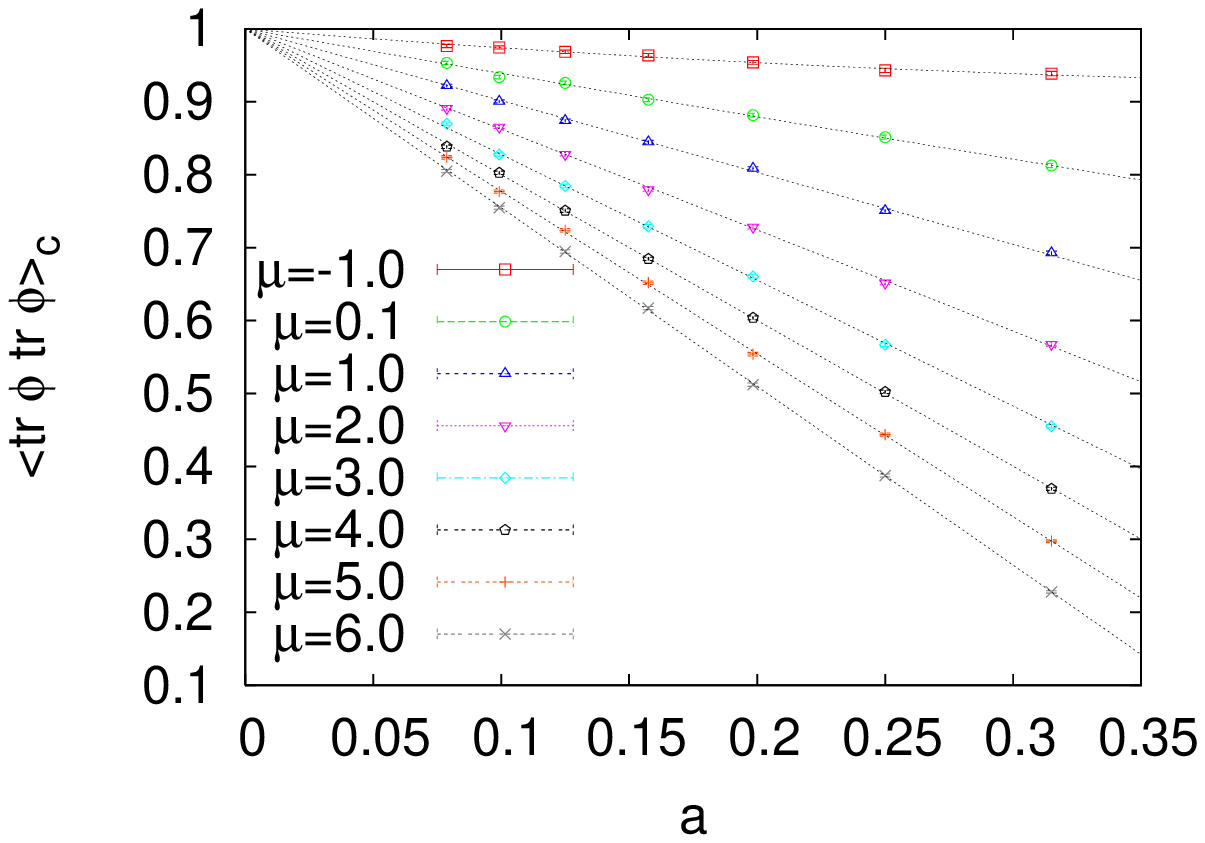, width=.49\textwidth}
{The observable $\langle \tr \phi^2 \, \tr \phi^2 \rangle_{\rm c}$ 
is plotted against $a(\equiv N^{-1/3})$ for various $\mu$
with $N=32,64,\cdots , 2048$.
For each $\mu$ we fit the data
to the behavior (\ref{O_p2p2})
without the ${\rm O}(a^4)$ terms
treating $h(\mu)$ as a fitting parameter.
\label{dsl_graph2}}
{The observable $\langle \tr \phi \, \tr \phi \rangle_{\rm c}$ 
is plotted against $a(\equiv N^{-1/3})$ for various $\mu$
with $N=32,64,\cdots , 2048$.
For each $\mu$ we fit the data
to the behavior (\ref{O_pp})
without the ${\rm O}(a^3)$ terms
treating $h(\mu)$ as a fitting parameter.
\label{dsl_graph1}}

\section{Next-leading $1/N$ corrections}
\label{section:exact}

So far we have been analyzing our Monte Carlo data
without using the knowledge obtained from analytical
results. 
The purpose of this section is to discuss
more detailed behaviors in the double scaling limit
which are obtained analytically, and to see whether
our Monte Carlo data reproduce those behaviors as well.

As we briefly review in the Appendix,
one can actually derive the asymptotic large-$N$ 
behavior of the correlation functions (for even $N$) 
in the double scaling limit as
\bear
\langle \tr \phi^2 \, \tr \phi^2 \rangle_{\rm c} &=&
2-\Bigl\{ \mu+h^2(\mu)\Bigr\} a^2 -\frac{1}{2}\Bigl\{
\mu \, h(\mu)-h^3(\mu)\Bigr\}a^3 +  {\rm O}(a^4) \ ,
\label{O_p2p2}\\
\langle \tr \phi \, \tr \phi \rangle_{\rm c} &=& 1-h(\mu)\, a-
\frac{1}{4}\Bigl\{ \mu-h^2(\mu)
\Bigr\} a^2 +
{\rm O}(a^3) \ ,
\label{O_pp}
\eear
where we have introduced a parameter $a\equiv N^{-1/3}$,
and $h(\mu)$ is a function which 
satisfies the differential equation \cite{Douglas:1990xv}
\be
\mu \, h(\mu) = h^3(\mu) -2 \, h''(\mu) \ ,
\label{PainleveII}
\ee
and the boundary conditions
%
\beq
h(\mu) \sim 
\left\{
\begin{array}{cc}
\sqrt{\mu} \quad & \textrm{for~} \mu\sim \infty \ , \\
0 \quad & \textrm{for~} \mu\sim -\infty \ . 
\end{array}
\right.
\label{bc2}
\eeq
Equation (\ref{PainleveII}) 
is nothing but the Painleve-II equation,
which is proven \cite{Hastings:1980} to 
have a unique real solution\footnote{In 
the case of $\phi^3$ matrix model,
which corresponds to the noncritical string theory
without worldsheet supersymmetry, one can obtain
only one boundary condition, since one can approach
the critical point only from one direction.
Accordingly the solution of the Painleve equation
has a one-parameter ambiguity \cite{Douglas:1989ve}.
This is essentially
because the vacuum of the matrix model
is nonperturbatively
unstable. The ambiguity arises from how one regularizes the
instability. The model we study in this paper does not 
have this problem.} 
under the boundary conditions (\ref{bc2}).
The solution is obtained numerically in ref.\ \cite{num}
to high accuracy, and we use it in plotting the exact
results in figs.\ \ref{dsl_p2p2}, \ref{dsl_pp} 
and \ref{fig_h}.

{}From (\ref{O_p2p2}) and (\ref{O_pp}),
the large-$N$ limits of the quantities
(\ref{p2p2_extracted}), (\ref{pp_extracted})
are obtained as
\beqa
\lim _{N\to \infty}A(\mu, N) &=& \mu + h^2(\mu)  \ , 
\label{AtoN}
\\
\lim _{N\to \infty}B(\mu, N) &=& h(\mu)  \ , 
\label{BtoN}
\eeqa
which we plot as exact results in figs.\ 
\ref{dsl_p2p2} and \ref{dsl_pp}.
Plugging in the boundary conditions (\ref{bc2}),
we reproduce eqs.\ (\ref{AtoNlim}) and (\ref{BtoNlim}) 
obtained from the planar results.

The analysis in the previous section therefore amounts to
extracting the leading $1/N$ corrections in
(\ref{O_p2p2}) and (\ref{O_pp}).
The reason for the observed slow approach to the 
limit (\ref{BtoN}) for $\mu \lesssim 0$ is that
the coefficient of the O($a$) term in the expansion (\ref{O_pp})
becomes much smaller than that of the O($a^2$) term
as $\mu$ decreases due to the boundary conditions (\ref{bc2}).

\FIGURE[t]{
\epsfig{file=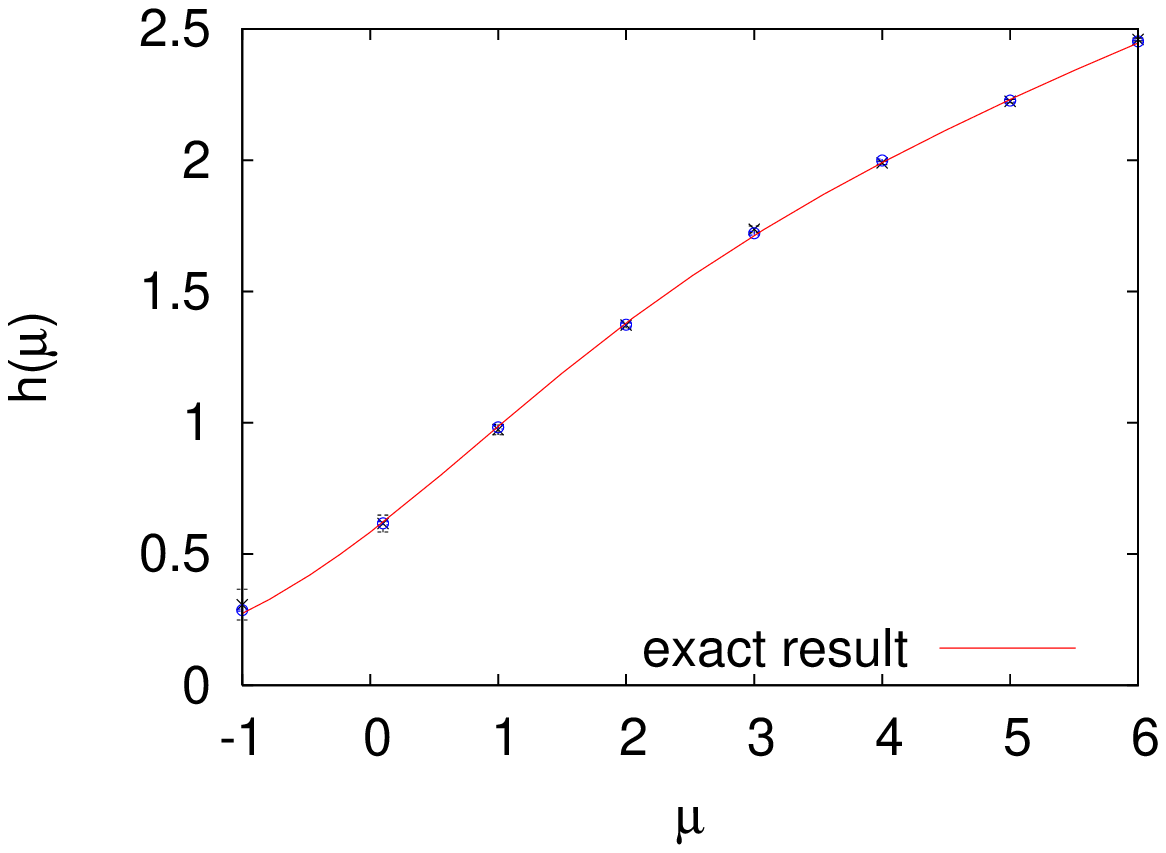, width=.5\textwidth}
\caption{
The crosses and the circles represent
the function $h(\mu)$, which is 
extracted from figs.\ \ref{dsl_graph2} and \ref{dsl_graph1},
respectively.
The solid line represents
the solution of the Painleve-II equation.
} 
\label{fig_h}
}

It would therefore be interesting to see whether
the {\em next-leading} $1/N$ corrections in
eqs.\ (\ref{O_p2p2}) and (\ref{O_pp}) are reproduced
by Monte Carlo simulation.
In fig.\ \ref{dsl_graph2} 
we plot 
$\langle \tr \phi^2 \, \tr \phi^2 \rangle_{\rm c}$
against $a$ for various $\mu$.
Indeed the data can be nicely fitted
to the behavior (\ref{O_p2p2})
without the ${\rm O}(a^4)$ terms,
where $h(\mu)$ is determined as a fitting parameter
by optimizing the fit for each $\mu$.
In fig.\ \ref{dsl_graph1}
we plot the observable 
$\langle \tr \phi \, \tr \phi \rangle_{\rm c}$ against $a$ 
for various $\mu$.
Again the data can be nicely fitted
to the behavior (\ref{O_pp})
without the ${\rm O}(a^3)$ terms,
where $h(\mu)$ is determined similarly.
The function $h(\mu)$ obtained in this way
is plotted in fig.\ \ref{fig_h}.
The crosses and the circles represent the results
obtained from 
$\langle \tr \phi^2 \, \tr \phi^2 \rangle_{\rm c}$
and $\langle \tr \phi \, \tr \phi \rangle_{\rm c}$, respectively,
which turn out to be consistent with each other
within error bars. 
Furthermore the results agree with 
the solution of the Painleve-II equation (\ref{PainleveII})
with the boundary conditions (\ref{bc2}).

\section{Summary}
\label{section:summary}

In this paper we have shown how one can use
Monte Carlo simulation to search for a double scaling
limit, and, if it exists, to obtain the corresponding
scaling functions.
For that purpose we studied a solvable one-matrix model
which has recently been proposed as a constructive
formulation of noncritical strings with worldsheet
supersymmetry.
In particular, we have shown how the results in the 
planar limit provide useful information in such an
investigation.
The required matrix size is not very large in most cases,
but we have also encountered a case in which
the approach to the large-$N$ limit turns out to be 
slow due to large next-leading $1/N$ corrections.

Considering that even a simple two-matrix model
are not solvable except for some special cases 
\cite{Itzykson:1979fi},
we believe that Monte Carlo simulation provides
a powerful tool to investigate the universality class
of matrix models in the double scaling limit.
For instance, in ref.\ \cite{Fukuma:2006ny}
a string field theory of minimal $(p,q)$ superstrings
has been constructed from the two-cut ansatz for
the two-matrix model.
It would be interesting to confirm their results
by taking the double scaling limit explicitly.

In general, if there exists a continuous phase transition
in the planar limit,
one has a chance to take the double scaling limit by
approaching the critical point with increasing $N$.
How generically this holds needs to be investigated.
For instance, it is known that the 
unitary matrix model \cite{Gross:1980he} has
a third order phase transition, which allows 
a double scaling limit \cite{Periwal:1990gf}.
The obtained limit belongs to the same universality class
\cite{Douglas:1990xv} as the one studied in this paper.
Whether a double scaling limit defines a sensible 
nonperturbative string theory is also an important issue,
which was addressed in refs.\ 
\cite{Ambjorn:2000bf,Anagnostopoulos:2001cb,Hanada}
by Monte Carlo simulation.
We hope that Monte Carlo studies of matrix models
will also shed light on nonperturbative dynamics
of critical strings.

\acknowledgments

It is our pleasure to thank Takehiro Azuma,
Masanori Hanada and Hirotaka Irie 
for valuable discussions.



\appendix

\section{Derivation of the asymptotic behaviors 
(\ref{O_p2p2}) and (\ref{O_pp})}
\label{appendix:scaling}

The prediction for the present model is obtained by 
the orthogonal-polynomial technique,
which is a powerful tool to
calculate various quantities in the double scaling
limit (see \cite{DiFrancesco:1993nw} for a review).
In this Appendix we briefly review the derivation of 
the asymptotic behaviors (\ref{O_p2p2}) and (\ref{O_pp})
for the reader's convenience.


Using the orthogonal-polynomial method,
various quantities in the matrix model 
can be expressed in terms of the coefficients 
$R_n \ (n=1,2,3,\cdots)$ 
characterized by the recursion formula
\bear
g \, \frac{n}{N} &=& R_n \, (-2+R_{n+1}+R_n+R_{n-1}) \ . 
\label{R_eq1}
\eear
For example, the correlation functions defined by
eqs.\ (\ref{O_p2p2}) and (\ref{O_pp}) are expressed as
\bear
\langle \tr \phi^2 \, \tr \phi^2 \rangle_{\rm c} 
&=& R_N \, (R_{N+1}+R_{N-1}) \ ,
\label{o2} \\
\langle \tr \phi \, \tr \phi \rangle_{\rm c} &=& R_N \ .
\label{o1}
\eear

In the planar limit (i.e., the $N\rightarrow \infty$ limit
with fixed $g$),
the asymptotic behavior of the coefficients $R_n$
is given by
\bear
R_n=\left\{
\begin{array}{cc}
\frac{1}{3}\left(1+\sqrt{1+3 \xi}\right) 
\quad & \textrm{for } \xi \geq 1  \ , \\
1-(-1)^n \sqrt{1-\xi}
\quad & \textrm{for } \xi \leq 1  \ , \\
\end{array}
\right.  \label{R_largeN}
\eear 
where $\xi = g n/N$ is regarded as a continuous variable.
Note that, for $\xi\leq 1$, the asymptotic behavior 
of $R_n$ is given by two continuous functions depending 
on the parity of $n$.
By plugging (\ref{R_largeN}) into (\ref{o2}) and (\ref{o1}),
one obtains the planar results (\ref{p2p2_largeN})
and (\ref{pp_largeN}).

Next we consider the double scaling limit;
i.e., the $N\rightarrow \infty$ limit
with fixed $\mu$ defined by (\ref{mu}) with $p=2$.
This implies that the coupling constant $g$ approaches
the critical point $g_{\rm cr}\equiv 1$ as
\beq
g = 1 - \mu a^2 \ ,
\eeq
where we have defined
$a\equiv N^{-1/3}$ as before.
In order to obtain the asymptotic behaviors of 
(\ref{o2}) and (\ref{o1}) in that limit,
we need to know the behavior of the coefficient
$R_n$ for the region of $n$, which can be parametrized as
\beqa
\xi &=& g \, \{ 1- (t-\mu) a^2 \}   \nonumber \\
 & =& 1-t a^2 + {\rm O}(a^4)
\label{xi-def}
\eeqa
using the new variable $t$.
For large $|t|$, we can deduce the asymptotic behavior
of $R_n$ from the planar result (\ref{R_largeN}).
Namely, by plugging (\ref{xi-def}) into (\ref{R_largeN})
and by expanding it with respect to $a$, we obtain
\bear
R_n=\left\{
\begin{array}{cc}
1-\frac{t}{4} a^2+ {\rm O}(a^4)\quad & 
\textrm{for~} t \sim -\infty \ , \\
1-(-1)^n \sqrt{t}a+ {\rm O}(a^3)
\quad & \textrm{for~} t \sim \infty  \ .
\end{array}
\right. \label{sl1}
\eear 
This motivates us to adopt the Ansatz 
\cite{Crnkovic:1990mr,Douglas:1990xv}
\beq
R_n= 1-(-1)^n H(t)\, a +F(t)\, a^2\ , \label{ansatz}
\eeq
where $H(t)$ and $F(t)$ are 
regarded as continuous functions of $t$,
which can be expanded with respect to $a$ as
\beqa
H(t)&=& h(t)+ {\rm O}(a^2) \ , \\
F(t)&=& f(t)+ {\rm O}(a^2) \ .
\eeqa
Substituting the Ansatz (\ref{ansatz}) into (\ref{R_eq1}), 
we obtain 
\bear
h^{\prime\prime}(t)&=&2 \, f(t) \, h(t) \ ,\\
h^2(t)&=&4 \, f(t)+t 
\label{f-h-rel}
\eear
as consistency conditions.
Eliminating $f(t)$, we obtain 
the Painleve-II equation (\ref{PainleveII}).
The asymptotic behavior (\ref{sl1}) translates into
the boundary condition\footnote{This is 
analogous to the case of unitary matrix
model \cite{Crnkovic:1990ms}.}
(\ref{bc2}).
Plugging (\ref{ansatz}) 
into eqs.\ (\ref{o2}) and (\ref{o1}),
we obtain the asymptotic behaviors 
(\ref{O_p2p2}) and (\ref{O_pp}).


\begin{thebibliography}{99}

\bibitem{'tHooft:1973jz}
 G.~'t Hooft,
\emph{A planar diagram theory for strong interactions},
\npb{72}{1974}{461}.

\bibitem{Aharony:1999ti}
 O.~Aharony, S.~S.~Gubser, J.~M.~Maldacena, H.~Ooguri and Y.~Oz,
  \emph{Large N field theories, string theory and gravity},
  \prep{323}{2000}{183} [\hepth{9905111}].


\bibitem{Gopakumar:1998ki_Dijkgraaf:2002fc}
  R.~Gopakumar and C.~Vafa,
  \emph{On the gauge theory/geometry correspondence},
  \atmp{3}{1999}{1415} [\hepth{9811131}] ;
  R.~Dijkgraaf and C.~Vafa,
  \emph{Matrix models, topological strings, and 
  supersymmetric gauge theories},
  \npb{644}{2002}{3} [\hepth{0206255}].
 
\bibitem{Kontsevich:1992ti}
  M.~Kontsevich,
  \emph{Intersection theory on the moduli space of curves 
  and the matrix Airy function},
  \cmp{147}{1992}{1}.


\bibitem{Brezin:1990rb}
  E.~Brezin and V.~A.~Kazakov,
\emph{Exactly solvable field theories of closed strings},
\plb{236}{1990}{144}.

\bibitem{Douglas:1989ve}
M.~R.~Douglas and S.~H.~Shenker,
\emph{Strings in less than one-dimension},
\npb{335}{1990}{635}.

\bibitem{Gross:1989vs}
D.~J.~Gross and A.~A.~Migdal,
\emph{Nonperturbative two-dimensional quantum gravity},
\prl{64}{1990}{127}.


\bibitem{DSLgeneral}
G.~Bertoldi,
\emph{Double scaling limits and twisted non-critical superstrings},
\jhep{0607}{2006}{006} [\hepth{0603075}];
G.~Bertoldi, T.~J.~Hollowood and J.~L.~Miramontes,
\emph{Double scaling limits in gauge theories and matrix models},
\jhep{0606}{2006}{045} [\hepth{0603122}];
  M.~Alimohammadi and M.~Khorrami,
\emph{Phase transitions of large-N two-dimensional 
Yang-Mills and generalized Yang-Mills theories in the double scaling limit},
\epjc{47}{2006}{507} [\hepth{0604027}];
L.~Alvarez-Gaume, P.~Basu, M.~Marino and S.~R.~Wadia,
\emph{Blackhole/string transition for the small Schwarzschild blackhole 
of AdS$_5$ $\times$ S$^5$ and critical unitary matrix models},
\epjc{48}{2006}{647}
[\hepth{0605041}].

\bibitem{Bietenholz:2002ch}
  W.~Bietenholz, F.~Hofheinz and J.~Nishimura,
\emph{The renormalizability of 2D Yang-Mills theory on a non-commutative
geometry},
  \jhep{0209}{2002}{009} [\hepth{0203151}].


\bibitem{Bietenholz:2004xs}
  W.~Bietenholz, F.~Hofheinz and J.~Nishimura,
\emph{Phase diagram and dispersion relation 
of the non-commutative lambda $\phi^4$ model in d = 3},
\jhep{0406}{2004}{042} [\hepth{0404020}].

\bibitem{Bietenholz:2006cz}
  W.~Bietenholz, J.~Nishimura, Y.~Susaki and J.~Volkholz,
  \emph{A non-perturbative study of 4d U(1) 
non-commutative gauge theory: The fate of one-loop instability},
  \jhep{0610}{2006}{042} [\hepth{0608072}].


\bibitem{BFSS} T.~Banks, W.~Fischler, S.~H.~Shenker and L.~Susskind, 
\emph{M theory as a matrix model: A conjecture},
\prd{55}{1997}{5112} [\hepth{9610043}]. 

\bibitem{Ishibashi:1996xs}
N.~Ishibashi, H.~Kawai, Y.~Kitazawa and A.~Tsuchiya, 
\emph{A large-N reduced model as superstring},
\npb{498}{1997}{467} [\hepth{9612115}]. 

\bibitem{DVV} R.~Dijkgraaf, E.~Verlinde and H.~Verlinde, 
\emph{Matrix string theory},
\npb{500}{1997} 43 [\hepth{9703030}]. 

\bibitem{EK}
T.~Eguchi and H.~Kawai,
\emph{Reduction of dynamical degrees of freedom 
in the large N gauge theory},
\prl{48}{1982}{1063}.


\bibitem{Nakajima:1998vj}
  T.~Nakajima and J.~Nishimura,
\emph{Numerical study of the double scaling limit in two-dimensional 
large N reduced model},
\npb{528}{1998}{355} [\hepth{9802082}].

\bibitem{Gross:1980he}
D.~J.~Gross and E.~Witten,
\emph{Possible third order phase transition in the large N 
 Lattice gauge theory},
\prd{21}{1980}{446}.


\bibitem{GAO} A.\ Gonz\'{a}lez-Arroyo and M.\ Okawa,
{\it A twisted model for large $N$ lattice gauge theory},
\plb{120}{1983}{174};
{\it The twisted Eguchi-Kawai model: a reduced model for large N 
lattice gauge theory},
\prd{27}{1983}{2397}.\\

\bibitem{AMNS} J.\ Ambj\o rn, Y.M.\ Makeenko, J.\ Nishimura
and R.J.\ Szabo, {\it Finite N matrix models of noncommutative 
gauge theory}, \jhep{11}{1999}{029} [\hepth{9911041}];
{\it Nonperturbative dynamics of noncommutative gauge theory},
\plb{480}{2000}{399} [\hepth{0002158}];
{\it Lattice gauge fields and discrete noncommutative Yang-Mills 
theory}, \jhep{05}{2000}{023} [\hepth{0004147}].




\bibitem{MRS}
S.\ Minwalla, M.\ van Raamsdonk and N.\ Seiberg,
{\it Noncommutative perturbative dynamics},
\jhep{02}{2000}{020} [\hepth{9912072}].

\bibitem{Ambjorn:2000dx}
  J.~Ambjorn, K.~N.~Anagnostopoulos, W.~Bietenholz, T.~Hotta and J.~Nishimura,
\emph{Monte Carlo studies of the IIB matrix model at large N},
\jhep{0007}{2000}{011} [\hepth{0005147}].

\bibitem{Anagnostopoulos:2001yb}
  K.~N.~Anagnostopoulos and J.~Nishimura,
\emph{New approach to the complex-action problem and its application to a
nonperturbative study of superstring theory},
\prd{66}{2002}{106008} [\hepth{0108041}].

\bibitem{Ambjorn:2000bf}
  J.~Ambjorn, K.~N.~Anagnostopoulos, W.~Bietenholz, T.~Hotta and J.~Nishimura,
\emph{Large N dynamics of dimensionally reduced 4D SU(N) super Yang-Mills
theory},
\jhep{0007}{2000}{013} [\hepth{0003208}].



\bibitem{McGreevy:2003kb}
  J.~McGreevy and H.~L.~Verlinde,
\emph{Strings from tachyons: The c = 1 matrix reloaded},
\jhep{0312}{2003}{054} [\hepth{0304224}].

\bibitem{Klebanov:2003km}
  I.~R.~Klebanov, J.~M.~Maldacena and N.~Seiberg,
\emph{D-brane decay in two-dimensional string theory},
\jhep{0307}{2003}{045} [\hepth{0305159}].


\bibitem{Takayanagi:2003sm}
  T.~Takayanagi and N.~Toumbas,
  \emph{A matrix model dual of type 0B string theory in two dimensions},
  \jhep{0307}{2003}{064} [\hepth{0307083}].

\bibitem{Douglas:2003up}
  M.~R.~Douglas, I.~R.~Klebanov, D.~Kutasov, 
  J.~M.~Maldacena, E.~Martinec and N.~Seiberg,
  \emph{A new hat for the c = 1 matrix model},
  \hepth{0307195}.

\bibitem{Klebanov:2003wg}
  I.~R.~Klebanov, J.~M.~Maldacena and N.~Seiberg,
  \emph{Unitary and complex matrix models as 1-d type 0 strings},
  \cmp{252}{2004}{275} [\hepth{0309168}].

\bibitem{Cicuta:1986pu}
G.~M.~Cicuta, L.~Molinari and E.~Montaldi,
{\em Large N phase transitions in low dimensions},
\mpla{1}{1986}{125}.





\bibitem{Kawai:2004pj}
  H.~Kawai, T.~Kuroki and Y.~Matsuo,
\emph{Universality of nonperturbative effect in type 0 string theory},
\npb{711}{2005}{253} [\hepth{0412004}].



\bibitem{Hanada:2004im}
M.~Hanada, M.~Hayakawa, N.~Ishibashi, H.~Kawai, 
T.~Kuroki, Y.~Matsuo and T.~Tada,
\emph{Loops versus matrices: The nonperturbative aspects of noncritical
string},
\ptp{112}{2004}{131} [\hepth{0405076}].


\bibitem{Sato:2004tz}
  A.~Sato and A.~Tsuchiya,
\emph{ZZ brane amplitudes from matrix models},
\jhep{0502}{2005}{032} [\hepth{0412201}].


\bibitem{Ishibashi:2005zf}
  N.~Ishibashi, T.~Kuroki and A.~Yamaguchi,
\emph{Universality of nonperturbative effects in $c < 1$ 
noncritical string
theory},
\jhep{0509}{2005}{043} [\hepth{0507263}].

\bibitem{Matsuo:2005nw}
  Y.~Matsuo,
\emph{Nonperturbative effect in c = 1 noncritical string theory and Penner
model},
\npb{740}{2006}{222} [\hepth{0512176}].


\bibitem{deMelloKoch:2004en}
  R.~de Mello Koch, A.~Jevicki and J.~P.~Rodrigues,
\emph{Instantons in c = 0 CSFT},
\jhep{0504}{2005}{011} [\hepth{0412319}].

\bibitem{Ishibashi:2005dh}
  N.~Ishibashi and A.~Yamaguchi,
\emph{On the chemical potential of D-instantons in c = 0 
noncritical string theory},
\jhep{0506}{2005}{082} [\hepth{0503199}].

\bibitem{Fukuma:2005nm}
  M.~Fukuma, H.~Irie and S.~Seki,
\emph{Comments on the D-instanton calculus 
in (p,p+1) minimal string theory},
\npb{728}{2005}{67} [\hepth{0505253}].

\bibitem{Kuroki:2007an}
  T.~Kuroki and F.~Sugino,
\emph{T duality of the Zamolodchikov-Zamolodchikov brane},
\prd{75}{2007}{044008} [\hepth{0612042}].


\bibitem{Martin}
  X.~Martin,
\emph{A matrix phase for the $\phi^4$ scalar field on the fuzzy sphere},
\jhep{0404}{2004}{077} [\hepth{0402230}];
%
F.~Garcia Flores, D.~O'Connor and X.~Martin,
\emph{Simulating the scalar field on the fuzzy sphere},
PoS {\bf LAT2005} (2006) 262 [\heplat{0601012}].

\bibitem{Panero}
   M.~Panero,
\emph{Numerical simulations of a non-commutative theory: 
The scalar model on the fuzzy sphere}, \hepth{0608202};
%
\emph{Quantum field theory in a non-commutative space: Theoretical
predictions and numerical results on the fuzzy sphere},
SIGMA 2 (2006), 081 [\hepth{0609205}].


\bibitem{Brezin:1977sv}
E.~Brezin, C.~Itzykson, G.~Parisi and J.~B.~Zuber,
{\em Planar diagrams},
\cmp{59}{1978}{35}.


\bibitem{Douglas:1990xv}
  M.~R.~Douglas, N.~Seiberg and S.~H.~Shenker,
  \emph{Flow and instability in quantum gravity},
  \plb{244}{1990}{381}.



\bibitem{Hastings:1980}
S.~Hastings and J.~McLeod, 
\emph{A boundary value problem associated with the second Painleve
transcendent and the Korteweg de Vries equation}, 
\emph{Arch.\ Rational Mech.\ Anal.} {\bf 73} (1980) 31.



\bibitem{num}
M.~Praehofer and H.~Spohn,
\emph{Exact scaling functions for one-dimensional stationary KPZ growth},
\emph{J.\ Stat.\ Phys.} {\bf 115(1-2)} (2004) 255
[\condmat{0212519}].


\bibitem{Itzykson:1979fi}
  C.~Itzykson and J.~B.~Zuber,
\emph{The planar approximation. 2},
\jmp{21}{1980}{411}.


\bibitem{Fukuma:2006ny}
  M.~Fukuma and H.~Irie,
\emph{A string field theoretical description 
of (p,q) minimal superstrings},
\jhep{0701}{2007}{037} [\hepth{0611045}].


\bibitem{Periwal:1990gf}
  V.~Periwal and D.~Shevitz,
\emph{Unitary matrix models as exactly 
solvable string theories},
\prl{64}{1990}{1326}.


\bibitem{Anagnostopoulos:2001cb}
  K.~N.~Anagnostopoulos, W.~Bietenholz and J.~Nishimura,
\emph{The area law in matrix models for large N QCD strings},
\ijmpc{13}{2002}{555} [\heplat{0112035}].

\bibitem{Hanada}
  M.~Hanada, H.~Kawai, T.~Kanai and F.~Kubo,
\emph{Phase structure of the large-N reduced gauge theory and generalized
Weingarten model},
\ptp{115}{2006}{1167} [\hepth{0604065}];
%
  M.~Hanada and F.~Kubo,
\emph{String tension and string susceptibility in two-dimensional generalized
Weingarten model}, \hepth{0611207}.



\bibitem{DiFrancesco:1993nw}
  P.~Di Francesco, P.~H.~Ginsparg and J.~Zinn-Justin,
  \emph{2-D Gravity and random matrices},
  \prep{254}{1995}{1} [\hepth{9306153}].


\bibitem{Crnkovic:1990mr}
  C.~Crnkovic and G.~W.~Moore,
  \emph{Multicritical multicut matrix models},
  \plb{257}{1991}{322}.



\bibitem{Crnkovic:1990ms}
  C.~Crnkovic, M.~R.~Douglas and G.~W.~Moore,
  \emph{Physical solutions for unitary matrix models},
  \npb{360}{1991}{507}.

\end{thebibliography}
\end{document}